\documentclass[11pt]{elsarticle}

\pdfoutput=1

\makeatletter
\def\ps@pprintTitle{%
  \let\@oddhead\@empty
  \let\@evenhead\@empty
  \let\@oddfoot\@empty
  \let\@evenfoot\@oddfoot
}
\makeatother

\usepackage{url}
\usepackage{breakurl}
\usepackage[breaklinks,
            colorlinks = true,
            linkcolor = blue,
            urlcolor  = blue,
            citecolor = blue,
            anchorcolor = blue]{hyperref}

\usepackage{lineno}

\usepackage{graphicx}
\usepackage[margin=1.25in]{geometry}
\usepackage[usenames,dvipsnames]{color}

\newcommand\acp{\begin{center}
\rule[-0.2in]{\hsize}{0.01in}\\\rule{\hsize}{0.01in}\\
\vskip 0.1in Submitted to the  Proceedings\\ 
of the African Conference on Fundamental and Applied Physics
    \vskip 0.05in
    {\it Second Edition, ACP2021, March 7--11, 2022 --- Virtual Event}\\
\rule{\hsize}{0.01in}\\\rule[+0.2in]{\hsize}{0.01in} \\
\end{center}}

\usepackage[firstpage=true]{background}
\backgroundsetup{contents={\parbox{6.5in}{\acp}}, scale=1,placement=top,opacity=1,color=black,position={3.25in,1.2in}}

\usepackage{fancyhdr}
\fancypagestyle{plain}{%
  \fancyhf{}%
  \fancyhead[C]{}
  \fancyfoot[C]{\thepage}
}

\fancypagestyle{empty}{%
  \fancyhf{}%
  \fancyhead[C]{{\it ACP2021, March 7--11, 2022 --- Virtual Edition}}
  \fancyfoot[C]{\thepage}
}
\begin{document}

\begin{frontmatter}


\title{ASFAP summary report: Particles and their Applications}

\author[add1]{Mar\'ia Moreno Ll\'acer\corref{cor1}}
\ead{maria.moreno.llacer@cern.ch}
\cortext[cor1]{Corresponding Author}
\address[add1]{Instituto de F\'isica Corpuscular (IFIC), Centro Mixto Universidad de Valencia - CSIC, Valencia, Spain}

\begin{abstract}
\noindent 
The Second Biennal African Conference on Fundamental and Applied Physics (ACP) took place in March 2022. The scientific program included three summary reports on the African Strategy for Fundamental and Applied Physics (ASFAP). Here, the summary report of ``Particle Physics and related applications'' groups (Particle Physics, Nuclear Physics, Medical Physics, Astrophysics and Cosmology, Fluid and Plasma, and Complex Systems; and the cross-cutting fields: Accelerators, Instrumentation and Detectors, Computing, and Nuclear Energy) is presented. The scope of each of the groups, the events organised so far and their upcoming activities are summarised.
\end{abstract}


\end{frontmatter}





\section{Introduction}
\label{sec:intro}
\noindent
The scientific program of the Second Biennal African Conference on Fundamental and Applied Physics (ACP2021), that took place online in March 2022~\cite{acp2021,ACP2021-report}, included three working sessions with three summary reports on the African Strategy for Fundamental and Applied Physics (ASFAP)~\cite{asfap}. ASFAP effort started in 2019, with the objetive to develop a strategy to increase African education and research capabilities; improve collaborations; and inform policymakers, stakeholders and international partners on the strategic directions likely to impact African advancement. For the development of such strategy, input was requested to the whole community (from all career stages and nationalities) as letters of interest (LOI). ASFAP has 16 Physics Working Groups and 6 Engagement Groups. Here, the summary of the report of the ``Particle Physics and related applications'' groups is presented. At the time of ACP2021, around 25-30 LOIs were submitted by these groups.
 
\section{Particle Physics}
\label{sec:particlephysics}
\noindent

The scope of the Particle Physics group is to:
\begin{itemize}
\setlength\itemsep{0em}
\item Contribute to build a network of Particle Physicists in Africa.
\item Push forward the ongoing activities and foster cooperations between African researchers for both Experimental and Theoretical physics.
\item Address the possibilities of evolution and expansion of these involvements and drive future endeavors.
\item Collect scientific inputs from the African Particle Physics community (in form of written LOIs) from subgroups, to provide a shared roadmap for the field (white paper).
\end{itemize}

The research experience of the groups in Africa involved in Particle Physics is very broad. On theoretical physics, there are groups working on radiative corrections in the Standard Model and beyond, perturbative unitarity and boundedness from below for scalar potentials and electroweak precision tests, Naturalness and Veltman conditions, Charged Higgs phenomenology and Vector like quarks. On the experimental physics side, there is expertise on physics analysis, beam tests, electronics development and remote operations.

Given the vast amount of topics covered by the group, four subgroups have been proposed (with two conveners for each, an experimentalist and a theorist):
\begin{enumerate}
\setlength\itemsep{0em}
\item ``Fundamental constituents and forces'' (Higgs physics; electroweak and beyond Standard Model physics; direct searches)
\item ``Symmetries and composite structures'' (flavour physics, charge-parity violation; strong interaction, hadron physics, heavy ions; indirect searches; neutron electric dipole moment)
\item ``Light messengers'' (neutrino physics: neutrino parameters, charge-parity violation, beyond Standard Model)
\item ``Infrastructures''
\end{enumerate}

\begin{figure}[!htbp]
\begin{center}
\includegraphics[width=0.22\textwidth]{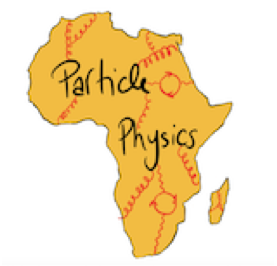}
\includegraphics[width=0.4\textwidth]{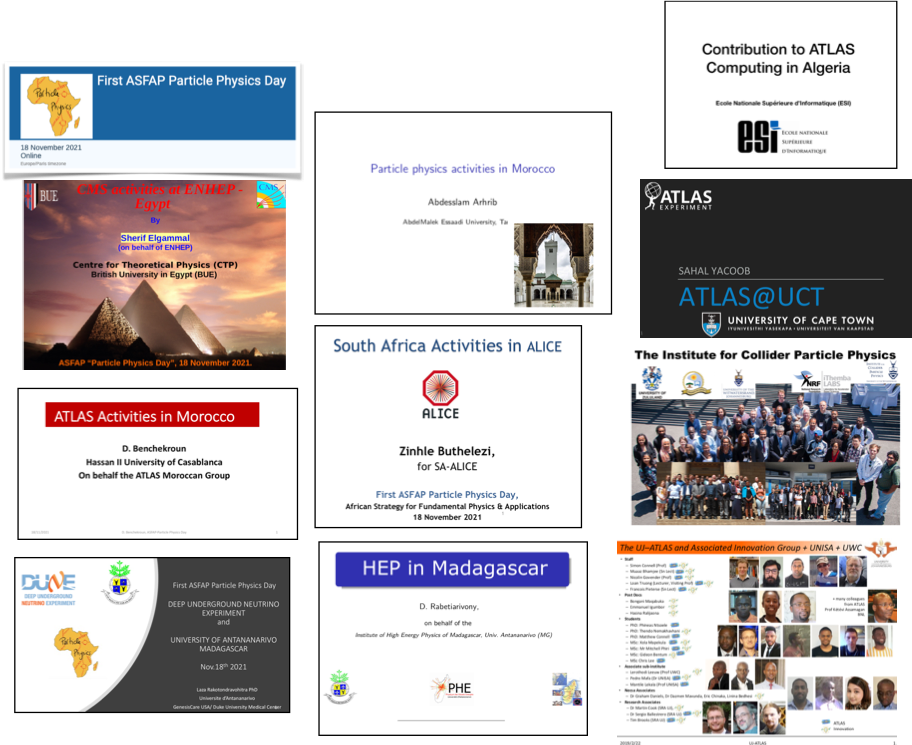}
\end{center}
\caption{Covers of some of the presentations of XXX}
\label{fig:particlephysics_collage}
\end{figure}

The Particle Physics group organised the ``First ASFAP Particle Physics Day'' in November 2021 with more than ten contributions from different groups around the whole African continent. It was an online event and it included reports from institutes working on the different experiments (ATLAS~\cite{ATLAS}, CMS~\cite{CMS}, ALICE~\cite{ALICE} and LHCb~\cite{LHCb}) of the Large Hadron Collider~\cite{LHC} at CERN, as well as reports from DUNE experiment on neutrino physics and theory groups, as shown in Figure~\ref{fig:particlephysics_collage}. The next event is planned for Spring 2022: ``the ASFAP Particle Physics PhD and postdocs day''. So far, 7 LOIs have been received.

The presence at CERN of the Particle Physics group is quite relevant, with involvement in experiments either as full members or associate (ATLAS, CMS and ALICE); also training opportunities in LHCb and computing WLCG. During the conference, it was pointed out several times that collaborations at CERN are very open, and the international laboratory is welcoming further participation of African physicists and groups. CERN can offer exciting science and training opportunities in an international context. Such networking also within Africa would also be highly beneficial.

\section{Nuclear Physics}
\label{sec:nuclearphysics}
\noindent
The first mini-workshop organised by the Nuclear Physics group took place in March 2022 with four contributions: i) ``ASFAP introduction''; ii) ``Nuclear Physics Activities at BIUST''; iii) ``The Pan African Virtual Nuclear University''; and, iv) report from a student ``Tanzania: challenges facing nuclear physics research (lack of suitable and qualified personnel, laboratory equipment for nuclear research, etc.)''. It followed with a discussion session in which few relevant aspects were brought-up, such as the need for a training session on Geant4 program (a nuclear and particle physics simulation software).
So far, 4 LOIs have been received: three on experimental facilities and one about education and training.

\section{Nuclear Energy}
\label{sec:nuclearenergyphysics}
The energy situation, which is very worrying in Africa, is evidenced by the absence, in most countries, of energy policies based on energy development master plans in the short, medium and long term. The ``Energy'' working group includes energy for a broad range of needs in Africa. It covered nine topics, ``Nuclear Energy'' is one of them. Such subgroup includes nuclear reactors, micro reactors, civil nuclear, synchrotron and fusion physics.

\noindent
The objectives of the Energy working group are to:
\begin{itemize}
\setlength\itemsep{0em}
\item{Strengthen cooperation/relations between African researchers and other actors in energy.}
\item{Identify scientific communities working in the field of energy, energy efficiency and sustainable development.}
\item{Create a dynamic of exchange and sharing of knowledge and know-how between academics and experts.}
\item{Have an overview and mapping on the research and innovation in energy area around Africa.}
\item{Facilitate the collaboration between researchers working in same topics}.
\item{Facilitate and encourage researchers to participate in the ASFAP project in an active manner and submit LOIs.}
\end{itemize}

The first meeting of this working group took place in February 2022 with contributions on the different kind of energies and regions in Africa. A second event is planned and, so far, 16 LOIs have been received.

The concrete scope of the ``Nuclear Energy'' subgroup is:
\begin{itemize}
\setlength\itemsep{0em}
\item \emph{Nuclear reactors}: Define and locate the African position on research and future of nuclear energy
\item \emph{Micro-reactors}: Research on the application of nuclear energy in health and civil fields
\item \emph{Civil nuclear}: Training and development in nuclear energy
\item \emph{Synchrotron}: Training and development in synchrotron techniques
\item \emph{Fusion physics}: Sharing information+results with actors and researchers in the field
\end{itemize}

\section{Medical Physics}
\label{sec:medicalphysics}
\noindent
The Medical Physics group organised its first mini-workshop in December 2021 with an overview of ASFAP mission and contributions from the International Atomic Energy Agency (IAEA), the Federation of African Medical Physics Organizations (FAMPO), and the platfrom Global Health Catalysts (GHT). Now, the focus is on bringing all Medical Physics Organisations together through FAMPO. One of the next meetings will be devoted for a joint sesion of the Medical Physics, Nuclear Physics, Energy and Instrumentation groups.

\section{Astrophysics and Cosmology}
\label{sec:astroandcosmophysics}
\noindent
The first mini-workshop of the Astrophysics and Cosmology group took place in November 2021, with more than 110 people involved (people from ~21 countries in Africa and from ~14 countries in Europe, South America and Asia). This working group is subdivided into different subgroups: solar physics; solar system, planetary sciences, astrobiology; stellar astronomy; galactic and extragalactic astronomy; cosmology and gravitational astronomy; transients and pulsars; high-energy astrophysics and astro-particle physics; astronomical instrumentation and infrastructure; astronomical methods and data; ethno-archeoastronomy (cultural astronomy); and astronomy for development. Close to 65\% of the participants are men and 35\% women.

The conveners of the working group have presented ASFAP initiative at more than 20 international conferences, and also distributed information about it among different networks and e-mail lists such as the African Astronomical Society, the African Network of Women in Astronomy, and Astronomy in Africa.

The plans for the upcoming months are: i) to start with regular meetings under each subgroup; ii) discuss about received LoIs and promote submission of new ones so that all topics are covered; iii) start shaping the White Papers; iv) discuss connections with other ASFAP working groups.

During the ACP2021, the second call for applications (schoolarships) of the Pan African Planetary and Space Science Network (PAPSSN) was advertised.

\section{Fluid and Plasma}
\label{sec:fluidandplasma}
\noindent
The Fluid and Plasma Physics group presented its strategy and initiatives in the ASFAP Town Hall meeting from July 2021. The focus on the group is on: i) non-thermal plasma generation (affordable low-cost methods); ii) environmental applications; iii) water purification; iv) disinfection; v) plasma agriculture (plasma-activated water); vi) plasma medicine; and, vii) aerospace applications. The role of plasma science and technology in enabling/augmenting new processes and applications is really broad, but it also faces grand challenges (infrastructure development, local technologies, education, etc.). Additional person power in this working group would be very benefitial.

\section{Complex Systems}
\label{sec:complexsystems}
\noindent
The Complex Systems includes theoretical groups working on applied mechanics such as biomechanics, solid mechanics, micro- and nano-structures dynamics, dynamical systems, and computer modeling. The work strategy includes these work-packages: WP1) identify the working groups in Africa and African scientists in the field all over the world; WP2) send out invitations and keep advertising to embark more people; WP3) form workgroups, assign tasks and come out with an activity timetable. Person power in the group is quite limited, so additional contributions would be very benefitial.

\section{Instrumentation and Detectors}
\label{sec:instrumanddetectors}
\noindent
The Instrumentation and Detectors Physics Group aims at promoting physics, development, design, implementation and evolution for a broad range of instrumentation and detectors applications in Africa. It is a transversal and multi-disciplinary group related to all physics groups. The goals of the group are: i) provide a coherent/flexible framework for efforts in instrumentation and detectors across Africa; ii) target groups: laboratories/centers, universities, pre-college; ii) bottom-up approach: individual and/or group collaborations; iv) encourage and help to submit LOIs; v) identify projects in instrumentation across countries, like shared facilities; vi) useful and meaningful efforts with tangible results; vii) cost effective implementation.

The first meeting of this working group took place in November 2021, with the goal to help the submission of LOIs by collecting and structuring information on existing facilities and projects in instrumentation. The main purpose of the meeting was to identify the most important future needs with respect to instrumentation in all fields of physics in Africa.

From a survey and listening to talks at ACP2021 and other meetings, it was concluded that the main problem that experimental researchers are facing is the need for experimental facilities and educational training centres in instrumentation for basic and applied experimental physics. 

Starting from existing LOIs and discussions at ACP2021, the conveners of the group have approached the LOI's authors directly and encouraged concretisation on the plans. They also discussed a very interesting proposal for a ``International Centre for Experimental Physics in Africa (ICEPA)'', common LOI of Instrumental and Physics Education working group. The idea is inspired by African Institute for Mathematical Sciences (AIMS) and other educational centres like Southern African Institute for Nuclear Technology and Sciences (SAINT). It will consist on a master-like curriculum typically one and a half year, including 6-month research project. It will include high-level lectures ($\leq$50\%) and hands-on experiences ($\geq$50\%), with a final examination and recognised diploma (association to university required in such case). The proposed educational training centre is similar to AIMS, but for experimental physics, strongly oriented towards instrumentation. Experimental installations and/or facilities will be installed at such a centre.

\section{Computing and $4^{\textnormal{th}}$ Industrial Revolution}
\label{sec:computing}
\noindent
The Computing and 4$^{\textnormal{th}}$ Industrial Revolution group is also a transversal one. In order to identify the needs for computing in the science community a survey was opened. The group covers a abroad range of topics: data science, (deep) machine learning, artificial intelligence, quantum science and technology, distributed computing, distributed analysis, computational physics, African infrastructure for networks and cloud services, e-learning, high performance computing, open source software, internet and bandwidth capacities, data centers, governance and policy, etc.
This working group is co-organizing ``International Conference on Data Science, Machine Learning and Artificial Intelligence'' conference, in collaboration with Namibia University of Science and Technology (NUST). The current status of this working group is described in Ref.~\cite{ASFAP_computing}.

\section{Accelerators}
\label{sec:accelerators}
\noindent
Accelerators is also a transversal working group, in connection with other groups such as Particle Physics, Nuclear Physics, Medical Physics and Radiation Biophysics, Material Science, Astrophysics and Cosmology, and Instrumentation and Detectors. The conveners started collecting a list of all existing facilities in Africa. This was not that trivial, as some of them cannot be found online, and there is also lack of information about the research that is being done in those facilities.
Another challenge is the fact that the Accelerator community in the African continent is very small due to lack of facilities where young people can be trained. Thus, before thinking of building more facilities on the African continent, the African youth needs to be trained with necessary skills.
iThemba Laboratory for Accelerator Based Science (LABS) in South Africa has recently been appointed as a collaborating center for IAEA. They are planning to host a training based on Electrostatic Accelerators which will take place in December 2022 at iThemba LABS. The training will target the audience from facilities in Africa to address the lack of skills in operation and maintenance. This initiative is a crucial starting point on this collaboration which will have a huge impact on the field of particle accelerator in African continent.

\section{Conclusions}
\label{sec:conc}
The summary of the report of the Particle Physics group and related applications of ASFAP at ACP2021 is presented here. It would be great to profit from the momentum of the conference and carry on with the planned activities and submit more LOIs.
It is essential to establish connections among the different African groups, and also synergies among the different working groups are also very important. Joint meetings/workshops and even liasions among the different groups are encouraged. In case there is no national-level Physics Society, the idea of appointing national contact points could be considered.
If any institute or university wants to get involved in the experiments at CERN, please get in touch with the ASFAP Steering Commitee. CERN makes cooperation agreements with governments once there is a solid basis of scientific cooperation.

\section*{Acknowledgments}
The author would like to thank the ACP2021 Organizing Committee and the ASFAP Steering Committee for their initiative.


\section*{References}
\bibliographystyle{elsarticle-num}
\bibliography{myreferences} 

\end{document}